\documentclass[times, twocolumn, 10pt]{article}
\usepackage{fullpage,slashbox,subfigure,setspace,times,url}
\usepackage{graphics,graphicx}
\usepackage{amssymb,amsmath,amsbsy, amstext, amsthm}

\newenvironment{definition}[1][Definition]{\begin{trivlist}
\item[\hskip \labelsep {\bfseries #1}]}{\end{trivlist}}

\newenvironment{example}[1][Example]{\begin{trivlist}
\item[\hskip \labelsep {\bfseries #1}]}{\end{trivlist}}

\title{The Aligned-Coordinated Geographical Routing for Multihop
  Wireless Networks}

\author{Ke Liu and Nael Abu-Ghazaleh \\
Computer Science Dept., SUNY Binghamton \\
\url{{kliu, nael}@cs.binghamton.edu}}

\date{}

\begin{document}
\maketitle
\begin{abstract} 

The stateless, low overhead and distributed nature of the Geographic
routing protocols attract a lot of research attentions recently. Since
the geographic routing would face void problems, leading to
complementary routing such as perimeter routing which degrades the
performance of geographic routing, most research works are focus on
optimizing this complementary part of geographic routing to improve
it. The greedy forwarding part of geographic routing provides an
optimal routing performance in terms of path stretch. If the
geographic routing could adapt the greedy forwarding more, its
performance would be enhanced much more than to optimize the
complementary routing such as perimeter routings. Our work is the
first time to do so. The aligned physical coordinate is used to do the
greedy forwarding routing decision which would lead more greedy
forwarding. We evaluate our design to most geographic routing
protocols, showing it helps much and maintain the stateless nature of
geographic routing. 
\end{abstract}

\section{Introduction}\label{intro}

\section{Related work}~\label{related}

\section{Connectivity Sensitive Alignment}\label{design}

\subsection{Intuition}
\label{subsec:intuition}
As our observation, in a wireless network with random deployment, the
2 data forwarding pathes in 2 directions between a pair of nodes are
mostly not through the same nodes, using GPSR as routing
protocol. And mostly, if in one direction, the path consists of both
greedy forwarding phase and perimeter routing phase, in the other
direction with different set of forwarding nodes, the path may keep
just in greedy forwarding as an optimal one in term of number of hops
(the shortest path).

\begin{example}
\begin{figure}[h]
\centering
\subfigure[From $S$ to $D$ ]{
  \label{fig:int1}
  \includegraphics[width=0.15\textwidth,height=0.21\textwidth]{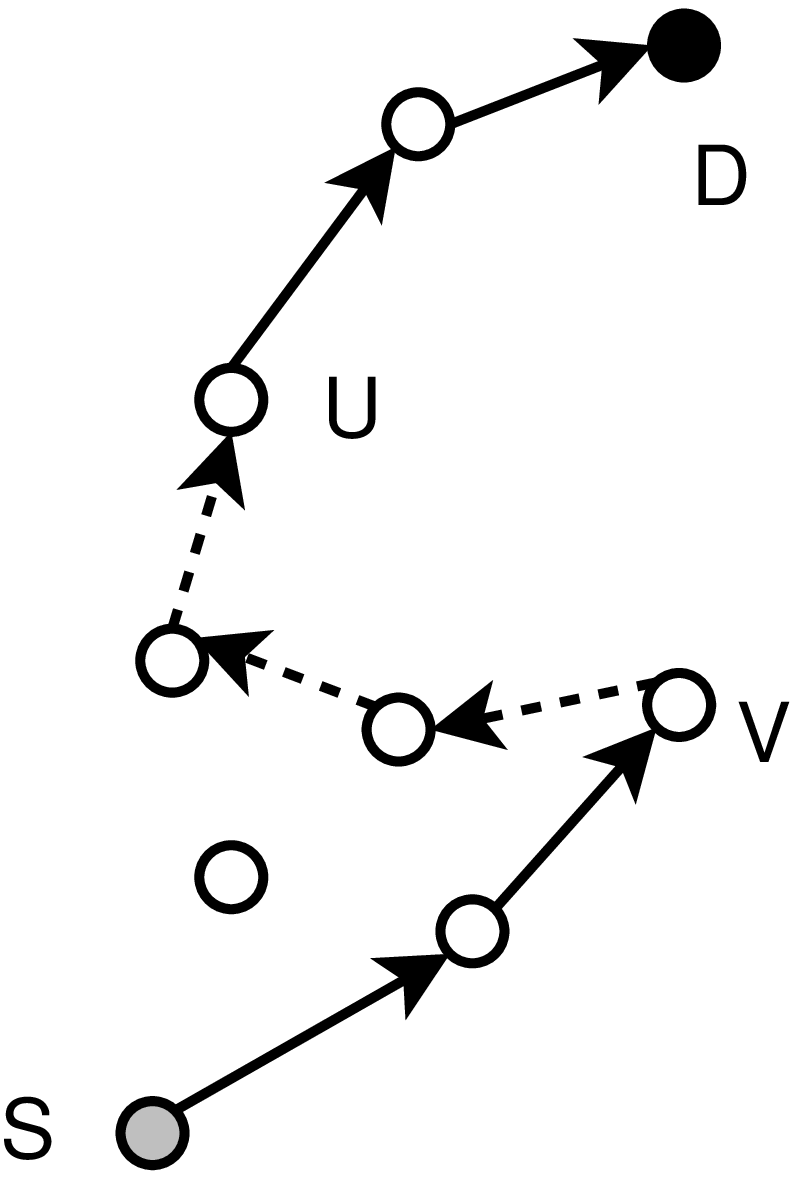}}
\hspace{.3in}
\subfigure[From $D$ to $S$]{
  \label{fig:int2}
  \includegraphics[width=0.15\textwidth,height=0.21\textwidth]{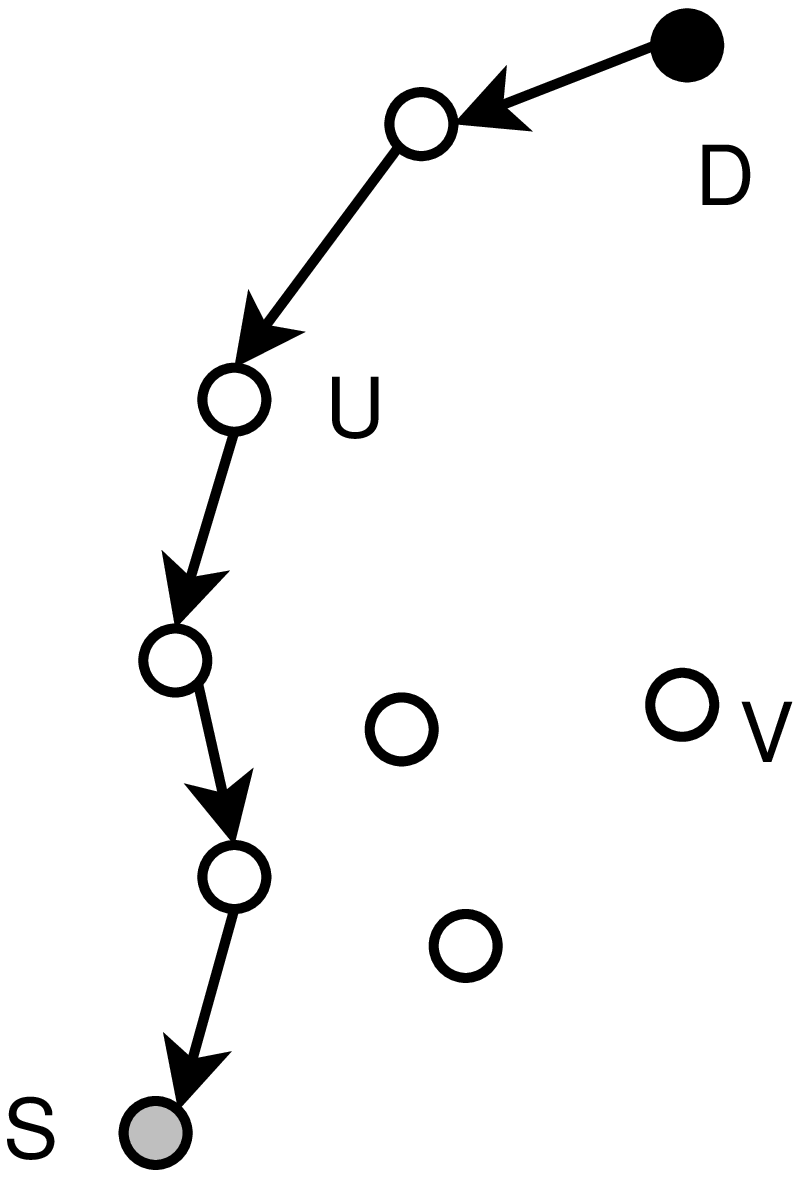}}
\caption{Communication pathes between node $S$ and $D$ through GPSR
  (concrete line denotes greedy forwarding, and dot line denotes
  perimeter routing.)}
\label{fig:intuition}
\end{figure}
The figure \ref{fig:intuition} shows a practice example of our
observation. From node $S$ to $D$, data packet would be forwarded
to node $V$ first, according to greedy principle; then it meets a
void, starting a perimeter routing phase; after reaching node $U$, it
comes back to greedy forwarding again, until arriving at destination
$D$, shown in figure \ref{fig:int1}.

Meanwhile, if a data packet is being forwarded from node $D$ to
$S$, the situation would be very different. It would be always
forwarded greedily, through a shorter path with 2 hops less than the
reverse one, as shown in figure \ref{fig:int2}.
\end{example}

Based on this observation, we questioned: is it possible to find a way
to reduce, or even extinct, the ratio of this situation so as to
increase the greedy ratio of the routing, resulting in an enhancement
of the routing performance.

Conservation routing protocols, such as DSR or AODV, are based on
broadcast and flooding. Before forwarding data packets, the routing
protocol would construct a shortest path between the source and
destination, with an overview of the whole network topology obtained
through the flooding. The overhead is high due to the flooding nature.
On the other hand, the stateless geogrpahic routing protocols such as
GPSR is based on only the location information of all one-hop
neighbors, leading to a lower overhead. The drawback is the void
problem where the greedy forwarding would fail, leading to
sub-optimal routing pathes.

In GPSR, the void may not be known before a packet reaches it.
The reason comes from the stateless nature of it, lack of overview of
not the whole network, but even 2 or more hops away.
If before facing a void, the routing protocol can find a way to predict
the void, and detour earlier, a perimeter routing phase is possible to
be avoided. Even worse, the one-hop information is not fully
utilized. Only the location information of them is used for routing.
Intuitively, if the connection information is used for routing as
well, it would help.

\subsection{Aligned coordinates of physical location}

The connectivity of a node is decided by all its neighbors, the
neighbors of neighbors, and so on. In this section, we try to align
the physical location of a node to all its neighbors, with the
connectivity information.

\begin{definition}
The aligned coordinates of a node, is a vector whose direction
and distance are decided by its neighborhood. Or say, the direction is
from its physical location to the average position of all its neighbors,
with a distance scaler as the standard deviation of the distances between
all its neighbors and itself.
\end{definition}

Suppose the location (physical coordinates) of node $X$ is $\langle
X\rangle$, the set of all its neighbor nodes is $N$, and $\langle
D_i\rangle$ is the location of the neighbor node $D_i$. The $|XD_i|$
denotes the distance between node $X$ and $D_i$, and
$\overline{\overline{N}}$ denotes the number of nodes in $N$. Then the
average position $\langle X_a\rangle$ of all neighbor nodes of $X$ is
\begin{equation}
\langle X_a \rangle = \overline{\langle D_i\rangle} =
\dfrac{\sum_{i\in N}{\langle D_i\rangle}}{\overline{\overline{N}}}
\label{eq:xa}
\end{equation}
The average distance $\overline{|XN|}$ between node $X$ and all its
neighbors is
\begin{equation}
\overline{|XN|} = \dfrac{\sum_{i\in
    N}{|XD_i|}}{\overline{\overline{N}}}
\label{eq:xn}
\end{equation}
The distance between the aligned location $X'$ and node $X$ meets the
standard deviation as
\begin{equation}
|XX'| = \sigma_X =
\frac{\sqrt{\sum{(|XD_i| - |XN|)^2}}}{\overline{\overline{N}}}
\label{eq:xx}
\end{equation}

So the aligned location $X'$ of node $X$ is
\begin{equation}
\langle X'\rangle = \widehat{\langle X_a\rangle}\cdot |XX'|
\label{eq:alignedx}
\end{equation}
where the $\widehat{\langle X_a \rangle}$ is the normalized vector of
$\langle X_a\rangle$.

As we can see, the alignment reflects the neighborhood connectivity of
a node with the standard deviation as the stretch. The average
location of all neighbors indicates to which direction, the node would
have a higher chance to find a neighbor, or next hop in routing. The
stretch of it shows how much the chance is.

\begin{definition}
The {\bf depth} of the alignment is the number of hops in which the
neighboring information is used for aligning.
\end{definition}
For example, if only the physical location information of all one-hop
neighbors is used for aligning of any node, the alignment depth of
this node is 1. If the aligned location information with depth $x$ of
all one-hop neighbors is used, the alignment depth of this node would
be $x+1$.

\begin{example}
\begin{figure}[h]
\centering
\includegraphics[width=0.3\textwidth]{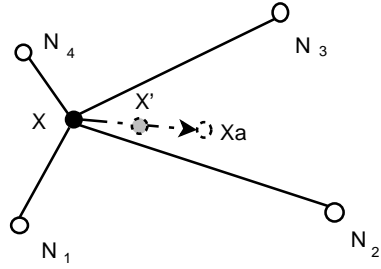}
\caption{The aligned location of a Node X}
\label{fig:avgpos}
{\scriptsize Xc is the average of neighbor nodes of X, and X'
  is aligned location of node X}
\end{figure}

The figure \ref{fig:avgpos} shows an example of how to align the
physical location of a node into the aligned coordinates.
According to equation \ref{eq:xa}, we have the average position
$\langle X_a\rangle$ of all neighbor nodes of $X$
\begin{equation*}
\langle X_a\rangle = \overline{\langle N_{i=1,2,3,4}\rangle} =
\dfrac{\sum_{i=1}^4{\langle N_i\rangle}}{4}
\end{equation*}
And the average distance between $X$ and its neighbor $N_i$s is
\begin{equation*}
|XN| = \frac{|XN_1| + |XN_2| + |XN_3| + |XN_4|}{4}
\end{equation*}
and the deviation is
\begin{equation*}
|XX'| = \frac{\sqrt{\sum_{i=1}^4(|XN_i|-|XN|)^2}}{4}
\end{equation*}
So we have the aligned coordinates with depth 1 of node $X$ as
\begin{equation*}
\langle X'_{(1)}\rangle=\overline{\langle N_{i=1,2,3,4}\rangle}\cdot
\frac{\sqrt{\sum_{i=1}^4(|XN_i|-|XN|)^2}}{4}
\end{equation*}
\end{example}

\subsection{Aligned Coordinates help Routing}

\begin{example}
\begin{figure}[h]
\centering
\includegraphics[width=0.35\textwidth]{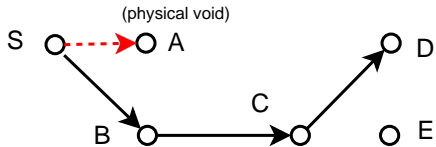}
\caption{Example: nodes in network}
{\scriptsize The dot line is the greedy forwarding path on physical
coordinates; The concrete line is the greedy forwarding path on
aligned coordinates.}
\label{fig:align-example}
\end{figure}

The figure \ref{fig:align-example} shows a 6-node network. Node $S$ is
the data source and the $D$ is the destination. Suppose the
radio range of the nodes is 1.5 unit. All other information is listed
in table \ref{tab:align-example}.

\begin{table}[h]
\centering \scriptsize
\begin{tabular}{|c|c|l|c|c|c|}
\hline X & $\langle X\rangle$ & $N$ & $\langle X'\rangle$
& $|XD|$ & $|X'D|$ \\
\hline S & $(0,1)$  & A, B    & $(0.31, 0.85)$ & $3$ & $2.70$\\
\hline A & $(1,1)$  & S, B    & $(0.81, 0.81)$ & $2$ & $2.20$\\
\hline B & $(1,0)$  & S, A, C & $(1.06, 0.28)$ & $2.24$ & $2.07$\\
\hline C & $(2.4,0)$& B, E    & $(2.35, 0.27)$ & $1.27$ & $0.98$ \\
\hline D & $(3, 0)$ & C, E    & $(2.88, 0.20)$ & - & - \\
\hline E & $(3, 0)$ & C, D    & $(2.91, 0.72)$ & $1$ & $0.81$\\
\hline
\end{tabular}
\caption{Routing Information for network in figure \ref{fig:align-example}}
\label{tab:align-example}
\end{table}
Since in a multihop wireless network, the alignment information of
several hop away may be difficult to obtained. Even obtained, it may
be stale quickly due to many reasons such as node mobility. To
maintain the stateless nature of geographic routing protocols, only
the un-aligned physical location information of the destination would
be used for distance calculating. So in stead of using $|X'D'|$,
$|X'D|$ is adapted.

As we can see, without alignment, at first data packet would be forwarded
greedily from node $S$ to node $A$. Since $|AD| < |BD|$, it reaches a
void, which may trigger a permeter routing. With alignment, $|A'D| >
|B'D|$, node $S$ would forward packet greedily to $B$, then $C$, and
finally $D$.
\end{example}

\section{Experiments}\label{sim}

\section{Conclusions and Future Work}~\label{conclude}

\bibliographystyle{unsrt}

\end{document}